\documentclass[]{interact}
\usepackage[caption=false]{subfig}

\usepackage[longnamesfirst,sort]{natbib}
\bibpunct[, ]{(}{)}{;}{a}{,}{,}
\renewcommand\bibfont{\fontsize{10}{12}\selectfont}

\theoremstyle{plain}

\theoremstyle{definition}

\theoremstyle{remark}

\begin{document}

\articletype{ARTICLE TEMPLATE}

\title{Leveraging Large Language Models for Enhancing Public Transit Services}

\author{
\name{Jiahao Wang\textsuperscript{a}\thanks{CONTACT Jiahao Wang. Email: jhope.wang@mail.utoronto.ca} and Amer Shalaby\textsuperscript{a}}
\affil{\textsuperscript{a}Department of Civil Engineering (Transportation Engineering), University of Toronto
}}
\maketitle

\begin{abstract}
Public transit systems play a crucial role in providing efficient and sustainable transportation options in urban areas. However, these systems face various challenges in meeting commuters' needs. On the other hand, despite the rapid development of Large Language Models (LLMs) worldwide, their integration into transit systems remains relatively unexplored.

The objective of this paper is to explore the utilization of LLMs in the public transit system, with a specific focus on improving the customers' experience and transit staff performance. We present a general framework for developing LLM applications in transit systems, wherein the LLM serves as the intermediary for information communication between natural language content and the resources within the database. In this context, the LLM serves a multifaceted role, including understanding users' requirements, retrieving data from the dataset in response to user queries, and tailoring the information to align with the users' specific needs. Three transit LLM applications are presented: Tweet Writer, Trip Advisor, and Policy Navigator. Tweet Writer automates updates to the transit system alerts on social media, Trip Advisor offers customized transit trip suggestions, and Policy Navigator provides clear and personalized answers to policy queries. Leveraging LLMs in these applications enhances seamless communication with their capabilities of understanding and generating human-like languages. With the help of these three LLM transit applications, transit system media personnel can provide system updates more efficiently, and customers can access travel information and policy answers in a more user-friendly manner.
\end{abstract}

\begin{keywords}
Public Transit System; Large Language Model; Customer Service; Intelligent Transit System
\end{keywords}

\section{Introduction}

Does this scenario sound familiar? Picture this: After an exhausting day at work, you find yourself at a bus stop, waiting for the bus to take you home. Predictably, yet still frustratingly, the bus is late. You resort to checking the transit agency's social media for real-time updates, which they claim to provide for your convenience. However, as is often the case, there's no update about your bus route. In response, you leave a comment criticizing not just the tardy bus, but also the lack of timely information. Later, back home and working on your weekend plans, you still consider using public transit despite the day's letdown. But as you start mapping out your journey, the complexity becomes apparent. The information overload is overwhelming: finding the most efficient connections, understanding the destination city's transit system, and ensuring your route is the best possible. The complexity of the information, coupled with your existing fatigue, tempts you to opt for a simpler bike ride around the city. Yet, you recall a recent experience when, after an hour-long bike ride, you were prohibited from bringing your bicycle onto the subway. A friend later informs you that bikes are allowed on buses. You hope that if only there were a more straightforward way to access such specific policy details from the transit agency, saving you time from navigating through extensive policy documents.

Public transit users encounter the same challenges mentioned earlier. Traditional transit systems often lack easy-accessed, personalized customer support for efficiently finding essential trip or system information. Offering personalized services, whether it's responding to customer requests, calls, feedback, or posting updates on social media, typically relies heavily on human staff. This human-centric approach significantly escalates operational costs for transit agencies. The clash between users' needs and the interests of transit providers underscores the pressing requirement for innovative solutions to enhance user experience in public transit while minimizing the additional burden on human workloads. 

To address these challenges, we propose a set of innovative and interconnected tools powered by Large Language Models (LLMs) to utilize the untapped potential of unstructured data in public transit systems by transforming it into easily digestible information for both the service provider and users. Our set of tools, collectively dubbed “TransitTalk”, are essentially digital assistants that facilitate the efficient dissemination of real-time system information and the enhancement of personalized interactions for transit users. For the service provider, our tools offer a solution for processing unstructured data – such as service updates – into easily understandable information to the transit users without heavily relying on the human workforce. This allows transit agencies to communicate effectively with passengers in real time, ensuring that they stay informed and can make necessary adjustments to their travel plans. For the public transit user, we provide interactive digital assistants, offering personalized trip guidance and relevant policy information tailored to the user’s needs. They can help users make informed decisions through an easy conversational format.

In this paper, we demonstrate how to use LLMs in the public transit domain with applications. Initially, we present a comprehensive information framework for constructing customer-interactive applications within the public transit domain, employing LLMs. This framework is designed to enhance customer interactions, enable personalized services, and elevate overall satisfaction. Then, we delve into the practical implementation of this framework through three specific applications: Tweets Writer, Trip Advisor, and Policy Navigator. These applications exemplify how the proposed structure can be effectively applied to address diverse transit-related needs and enhance user experiences.

The rest of this paper will be organized as follows. Section~\ref{sc:rw} provides a general introduction to LLMs, discussing their capabilities and advancements. Then, we explore LLM applications in non-transportation fields and transportation-related domains. Section~\ref{sc:method} presents the methodology and framework for using LLMs in public transit and example applications. Section~\ref{sc:discussion} discusses best practices, addresses limitations, and explores potential solutions. In the end, section~\ref{sc:con} presents possible future works.

\section{Related Work}\label{sc:rw}
\subsection{Overview of Large Language Models}

LLMs are advanced DL models utilized in Natural Language Processing (NLP) tasks, to comprehend and generate human-like texts. LLMs belong to the category of generative artificial intelligence (AI) models, capable of not only understanding and analyzing data but also generating novel outputs based on their training data. In this section, we provide a general overview of LLMs, including their key characteristics and different types.

\subsubsection{Key Characteristics}

\textbf{Transformer}: Transformer (\cite{vaswani2017attention}) is the foundation of LLMs. According to (\cite{vaswani2017attention}), there are multiple advantages of using a Transformer as the base model for LLMs, compared to other model structures that are widely used in Language Models, such as Recurrent Neural Networks (RNNs) and Convolutional Neural Networks (CNNs). First, Transformer has better performance in capturing long-range dependencies within input sequences. Transformer is easier to learn information not only in the sentence-dimensional, but also paragraph-dimensional, or even article-dimensional. Second, Transformer does not rely on the sequence of input series, which makes the model capable of parallel computing and can highly improve the model's training and computing efficiency. Last but not least, the structure of the Transformer, built up with Encoders and Decoders, is easy to modify, including the depth and width of the model. These features of Transformer make LLMs easy to have a deep model structure for long-text learning tasks and can be trained with relatively less time and computing resources (\cite{bietti2023birth}).

\textbf{Large}: Within the scope of Large Language Models (LLMs), the term 'Large' encapsulates two significant aspects: the expansive volume of training data and the massive scale of model parameters. Training an LLM is not solely reliant on a single book dataset; it harnesses the power of multifaceted resources ranging from online forums, Wikipedia, and news articles to code repositories. This diverse pool of training data provides LLMs with a holistic and macroscopic understanding of various linguistic styles, themes, and contexts, thereby enabling them to tackle a broader range of tasks (\cite{zhao2023survey, 2019t5}). Meanwhile, we observe a growing trend in the size of the model architecture, with parameter counts increasing into the billions, motivated by the requirement to optimally learn from increasing data sizes (\cite{brown2020language}). The growth in both the volume of training data and the number of model parameters endows LLMs with emergent capabilities, which are not shown in the smaller language models, such as in-context learning, instruction following, and step-by-step reasoning (\cite{wei2022emergent, zhao2023survey}).

\textbf{Training Process}: To address the challenge of training with extensive volumes of data, the LLM training pipeline is generally divided into two primary phases: Pre-training and Fine-tuning. The pre-training phase is the more time-consuming step, which involves self-supervised learning from large-scale datasets in a distributed manner. This phase enables the model to acquire a broad, general-purpose problem-solving capability (\cite{brown2020language}). On the other hand, the fine-tuning process sharpens the pre-trained model's skills towards specific tasks. This stage leverages relatively smaller computational resources and requires less training time, as it adjusts the model to support more task-orientated applications (\cite{zhao2023survey}). This structured training pipeline offers significant advantages to the end-users of LLMs. Users do not need to start the training process from scratch but can instead focus on tailoring the pre-trained model to their specific tasks.

\subsubsection{Types of Large Language Models}
Large Language Models (LLMs) primarily fall into two categories: Base Language Models and Instruction-Fine-tuned Language Models. Base Language Models, such as T5 (\cite{raffel2020exploring}), GPT-3 (\cite{brown2020language}), and PaLM (\cite{chowdhery2022palm}), are fundamental LLMs trained exclusively on large-scale text corpora without undergoing any fine-tuning processes for performance enhancement (\cite{korbak2023pretraining}). Trained to predict subsequent words in a given context, these models are not explicitly tailored for specific tasks but rather aim for the understanding of language structures and patterns. As for Instruction-Fine-tuned Language Models, such as T0 (\cite{sanh2021multitask}), GPT-3.5/4 (\cite{ouyang2022training}), and ChatGPT (\cite{roumeliotis2023chatgpt}), undergo an additional step of optimization based on base Language Models. Through fine-tuning with specific instructions, downstream tasks, or human feedback, these models gain the ability to deal with specific language-related tasks, including question answering, text classification, summarizing, writing, keyword extraction, etc (\cite{lou2023prompt}). This category of LLMs offers specialized performance, aiming for more task-focused applications of language models.

\subsection{Large Language Model Applications} 
\subsubsection{LLM Applications to Transportation Problems}
Researchers are increasingly incorporating LLMs into the development of solutions for conventional transportation tasks. For instance, in~\cite{ding2023exploratory}, LLMs were utilized to extract attributes from raw crash reports and organize the extracted data according to a pre-set format. Additionally, LLMs have shown promise in generating code for transportation-related software and data mining tasks within the transportation field. A demonstration of this application can be seen in~\cite{fu2023towards}, where an LLM was used to interpret urban information, generating code for ArcGIS and street view recognition programs. Furthermore, LLMs have been employed in sentiment analysis for analyzing citizens' behaviors. Beyond traditional information extraction and classification tasks that rely on text input, LLMs can also be used for cross-modal applications. These involve answering questions based on image or audio inputs. This type of application can be achieved by combining LLMs with other tools such as BLIP-2 (\cite{li2023blip}) for vision-to-language encoding, and Whisper (\cite{radford2023robust}) for audio recognition. One application of this approach, as demonstrated in~\cite{zheng2023chatgpt}, involved a cross-modal LLM answering questions based on images taken by bus cameras or audio recordings from car crash events. 

\subsubsection{Applications in Non-transportation Fields}

On the other hand, LLMs have demonstrated their features in diverse non-transportation fields, showing superior natural language understanding, generation, and manipulation capabilities. The adaptability of LLMs positions them as valuable tools with potential for adoption or adaptation in specific transportation contexts. Here, we explore how LLMs developed in different fields could find applications in transportation-related areas.

\textbf{Planning:}
The diverse applications of LLMs in fields such as optimizing construction timetable (\cite{prieto2023investigating}), supply chain decision-making (\cite{frederico2023chatgpt}), and marketing strategies (\cite{rivas2023marketing}) suggest promising contributions to transportation planning. LLMs could optimize transit driver schedules, enhance supply chain logistics for transportation infrastructure projects, and craft effective communication strategies for public transportation initiatives.

\textbf{Management:} In~\cite{liu2023radiology, singhal2022large}, LLMs were used to answer radiology image-related queries after being fine-tuned with structured datasets containing patients' imaging findings and corresponding interpretations. This offers potential applications in traffic management where we can give management suggestions for the given structured traffic situation information. Fine-tuning LLMs with structured data from on-/off-road sensors and corresponding traffic management strategies may empower them to analyze and respond to real-time traffic conditions, improving the efficiency of traffic flow and reducing congestion. The application of LLMs in multi-modal understanding and generation (\cite{wu2023next}) could further enrich the data sources, including traffic camera pictures, for fine-tuning LLMs in traffic management, and broaden the application scenarios.

\textbf{Education and Assessment:}
LLMs' proficiency in educational settings, as seen in the evaluation of ChatGPT by~\cite{kung2023performance}, suggests applications in transportation education and assessment. Leveraging Chain-of-Thought (\cite{lievin2022can}) or fine-tuning with domain-specific knowledge (\cite{liu2023context}) could enhance LLMs' performance in assessing the knowledge and competence of transportation professionals. Additionally, utilizing LLMs for teaching assistance tasks, such as text and exam content generation (\cite{kovavcevic2023use}) and feedback giving (\cite{rudolph2023chatgpt}), opens avenues for educational initiatives in the transportation sector.

\textbf{Human-Machine Interaction:}
The integration of LLMs into robotics suggests potential applications in human-autonomous vehicle interaction. Exemplified by~\cite{li2023stargazer}, a GPT-3 model was deployed to process human audio instructions, enabling the robot to move based on explicit movement directions and user preferences. In~\cite{chen2023open}, an LLM was employed to generate a single comprehensive plan for robot control. Moreover, in~\cite{wu2023tidybot}, the PaLM model provides summarization capabilities, helping a household robot understand the environmental images captured by its head camera, allowing it to recognize potential cleaning tasks and determine an effective sequence for their execution. Utilizing LLMs to process and understand human instructions can enhance communication between users and vehicles. This capability holds significant implications for the development of advanced driver assistance systems, enabling the integration of verbal commands, or may further shift the paradigm in how autonomous vehicles interpret and respond to human directives, contributing to safer and more intuitive interactions on the roads. 

\textbf{Research:}
LLMs have found applications in scientific research, serving as valuable tools for manuscript review (\cite{ufuk2023role}), research paper summarization, and generating recommendations for researchers (\cite{lee2020biobert}). The ability of LLMs to assist in writing academic papers (\cite{biswas2023chatgpt}) and even contribute as co-authors (\cite{siegerink2023chatgpt}) suggests their potential to ease the workload of researchers in the transportation domain. With the aforementioned abilities,  LLMs may assist in summarizing research papers, identifying key trends, and providing insights into the latest advancements and challenges in the transportation domain. Moreover, LLMs can enhance the data mining process for big data (\cite{hassani2023role}), offering valuable support for analyzing the growing amount of data collected in transportation systems.

In conclusion, the adaptability of LLMs presents exciting possibilities for addressing transportation challenges and enhancing various aspects of the sector. By considering the unique strengths of LLMs in specific transportation application areas, researchers and practitioners can unlock innovative solutions that contribute to the evolution of modern transportation systems.

\subsection{Discussion}
While reviewing the state-of-the-art LLM-based applications in the transportation field, we have noticed several limitations. First, these applications tend to focus more on demonstrating specific functions of the LLM rather than integrating it into real-world scenarios. Second, most of these applications are simple Q\&A systems, where there is only one-way communication with the LLM providing answers to user questions. Lastly, these applications lack sufficient decision-making resources as they heavily rely on user inputs without utilizing other supported resources for more reliable answers.

To address these limitations and specifically focus on the area of public transit, we introduce initial attempts in the following sections. We first introduce the basic methodologies, including Chain-of-Thought and vector database. Then, we propose the general structure for LLM applications in the transit domain. We further explore three guiding applications related to customer experience in the public transit system. Each application pertains to a type of information communication scenario that may arise. By showing the three guiding applications, we hope to give readers a better understanding of the technical requirements for building up similar transit digital assistants.

\section{Leveraging LLMs for Public Transit System Services}\label{sc:method}

Our objective in utilizing LLMs in the public transit system is to develop applications that enhance the customer experience by delivering real-time and personalized information to users. In this section, we begin by outlining the methods employed in the application development process, as well as the overarching framework that encompasses all three applications. Subsequently, we describe each application using hypothetical scenarios and provide insights into the underlying operational logic driving their functionality.

\subsection{Methodologies}
In building our applications, we employ a range of methodologies in leveraging LLMs for enhancing the public transit system experience, such as the Chain-of-Thought approach and Vector Database. Each of these methodologies contributes to the overall framework that underpins the three transit applications: Tweet Writer, Trip Advisor, and Policy Navigator. In the following subsections, we will delve into the specific details of each application, illustrating how these methodologies are applied to meet the technical requirements and drive the functionality of the three applications.

\subsubsection{Chain-of-Thought}
While LLMs are capable of generating answers to users' queries, it is important to note that the accuracy and consistency of these results is not guaranteed. According to a study by ~\cite{kojima2022large}, the accuracy of PaLM tested on MultiArith without any prompts (Zero-Shot) is significantly lower compared to cases using prompts. One particularly effective method for prompt engineering is the Chain-of-Thought (CoT) approach, initially proposed by ~\cite{wei2022chain}. The core idea behind CoT is to encourage the LLM to demonstrate the thought process it undergoes to arrive at the final answer, rather than simply providing a direct response.

\begin{figure}[!htbp]
  \centering
  \includegraphics[width=0.98\textwidth]{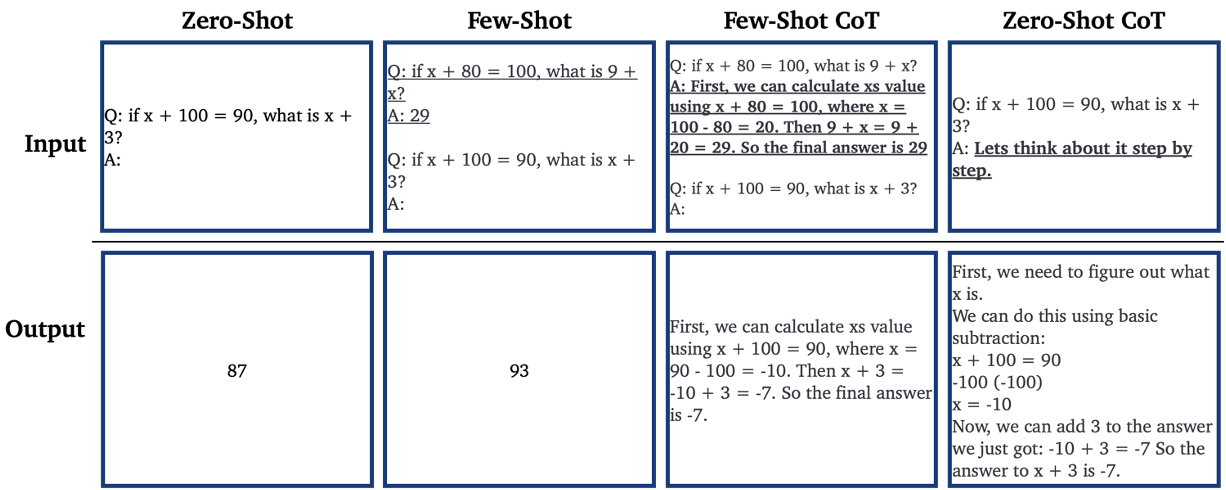}
  \caption{Comparison of the performance of LLM (text-davinci-003) solving a simple mathematical problem with different prompts.}\label{fig:cot}
\end{figure}

There are two main styles of CoT: Few-Shot CoT (\cite{wei2022chain}) and Zero-Shot CoT (\cite{kojima2022large}). In Few-Shot CoT, the LLM is provided with examples as references to guide the thought chain and derive the final result. Conversely, in Zero-Shot CoT, a simple yet effective approach involves adding the sentence "Let's think step by step" at the end of the original prompt. The experiment conducted in (\cite{kojima2022large}) demonstrates the performance improvement achieved through Zero-Shot CoT. Figure~\ref{fig:cot} 
illustrates the demonstration of both CoT methods, highlighting how employing CoT can enhance the likelihood of obtaining accurate answers from the LLM.

\subsubsection{Vector Database}
Another technology utilized in our application development is a vector database. This type of database is employed to address the requirements of querying (unstructured) resources with natural language input. Unlike traditional queries executed on structured databases, such as relational databases, it is not feasible to directly translate user queries into a specific database query language.

\begin{figure}[!htbp]
  \centering
  \includegraphics[width=0.65\textwidth]{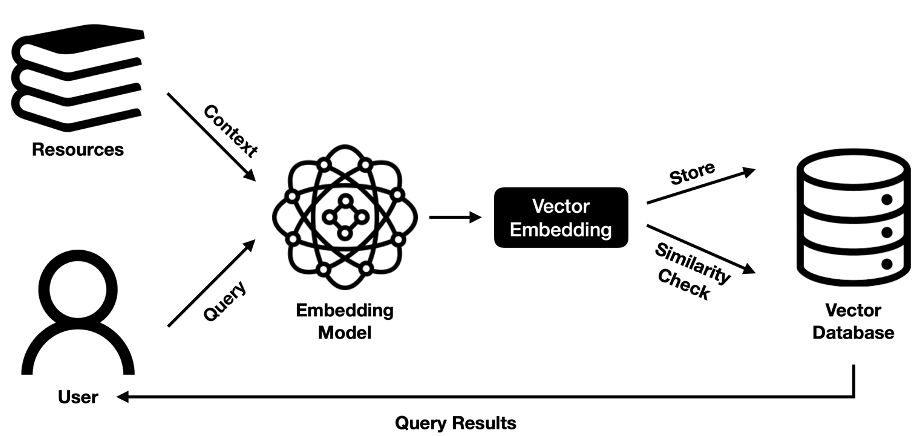}
  \caption{Demonstration of the workflow of vector database}\label{fig:vd}
\end{figure}

The concept of the vector database was initially introduced in~\cite{stata2000term}, where the fundamental workflow for storage and querying on the vector database is illustrated in Figure~\ref{fig:vd}. 
The primary function of the vector database lies in the embedding process for the context and user query. By utilizing vector databases, it becomes feasible to perform similarity searches, such as KNN, in high-dimensional vector spaces. This is achieved by converting the unstructured user query and context into fixed-length feature vectors. Notably, the integration of LLMs has significantly enhanced the effectiveness of embedding, resulting in more accurate information retrieval.

\subsubsection{General Application Structure}

\begin{figure}[!htbp]
\centering
\subfloat[Traditional Information System Structure in Transit System]{\label{fig:trad_frame}\includegraphics[width=0.45\textwidth]{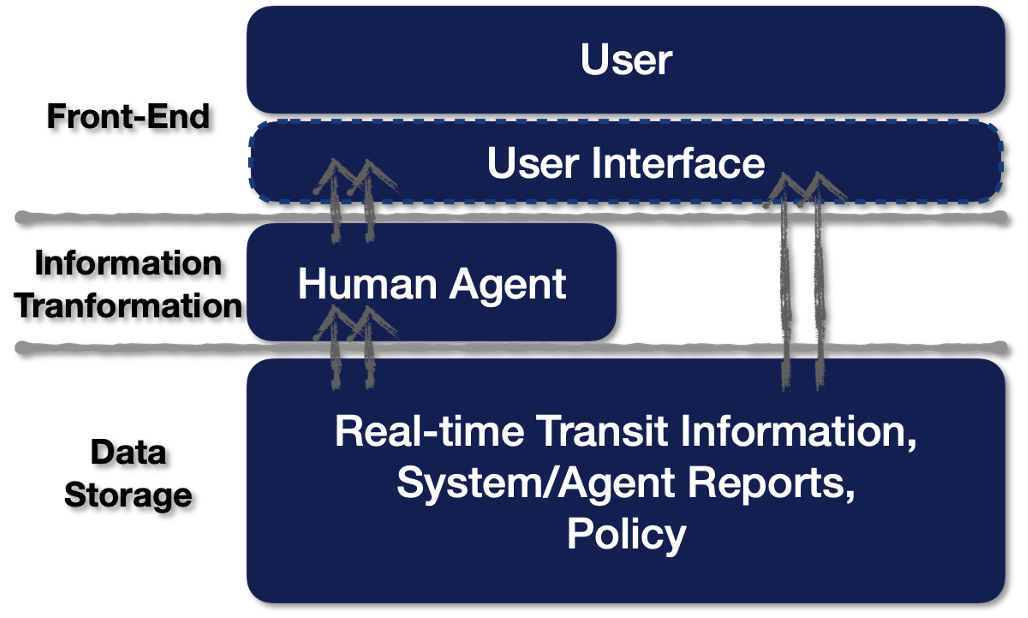}}
\hfill
\subfloat[Proposed Information System Structure in Transit System Embedding LLM]{\label{fig:gp}\includegraphics[width=0.45\textwidth]{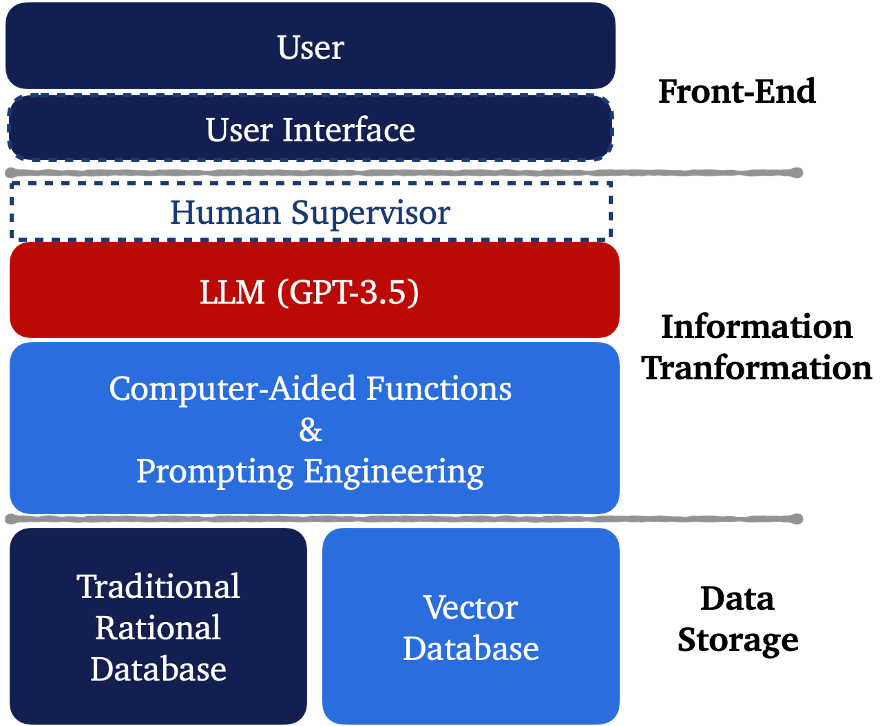}}
\caption{Comparison of The Traditional Information System Structure and The Proposed Structure for Public Transit System}\label{fig:animals}
\end{figure}

Before presenting the proposed application structure, let's begin by discussing the traditional information system structure. As depicted in Figure~\ref{fig:trad_frame}, data, encompassing structured information like system alerts and unstructured data such as policy information or feedback from agents, resides within the back-end data container. These data sets undergo processing by human agents before being disseminated to the front end. Alternatively, the data can be directly transmitted to the front-end application, where it is accessible to customers. 

However, during this process, human agents are confronted with a substantial volume of data, making it challenging to ensure the efficient publication of information. Conversely, providing all information without tailoring, such as policy details, can prolong the time it takes for customers to access valuable information and potentially diminish user satisfaction.

To optimize the information transformation process and deliver personalized information to transit users, we've integrated an Information System enhanced with Large Language Models (LLM), as depicted in Figure~\ref{fig:gp}. In this upgraded system architecture, we've enriched the information transformation layer with LLM, including its associated components such as computer-aided functions and prompting engineering.

Here's how it works: Raw data is processed using LLM, specifically GPT-3.5, and tailored to match user queries expressed in plain language. The tailored information with finally be presented to the user through the front-end user interface. Within this framework, the LLM assumes various roles, including natural language comprehension, information extraction, sentiment analysis, and summarization. This LLM-based approach empowers users to engage in conversational and automated interactions, making information retrieval more intuitive. Moreover, human agents are relieved from the manual task of converting raw transit system data. Instead, they oversee and ensure the accuracy and coherence of the information generated by the LLM.

However, it's important to note that the LLM alone may not handle the entire task. To guide the LLM in acquiring pertinent information for decision-making, we employ prompting engineering techniques, specifically in the form of CoT. Additionally, we integrate computer-aided functions to enable the LLM to access information through the transit agent's APIs. Furthermore, we implement a vector database within the application, facilitating natural language-based content queries for the LLM.

\subsection{Three LLM Applications to Transit}
In this subsection, we present the three transit applications utilizing LLMs:
\begin{itemize}
    \item \textbf{Tweet Writer}: to help the staff in the media department in transit systems to efficiently disseminate transit system alerts on social media platforms;
    \item \textbf{Trip Advisor}: to provide users with personalized trip recommendations;
    \item \textbf{Policy Navigator}: to offer clear and tailored responses to policy-related user queries.
\end{itemize}
All three applications require the LLM to provide accurate and consistent responses, adhering to specific formatting guidelines. These responses are constructed using information sourced from both structured and unstructured databases. By delving into these three applications, we gain a comprehensive understanding of the technical details of LLM applications, including the workflow and output examples, and see how can we ensure that LLM responses meet essential constraints and maintain accuracy and legitimacy.

\begin{figure}[!htbp]
  \centering
  \includegraphics[width=0.65\textwidth]{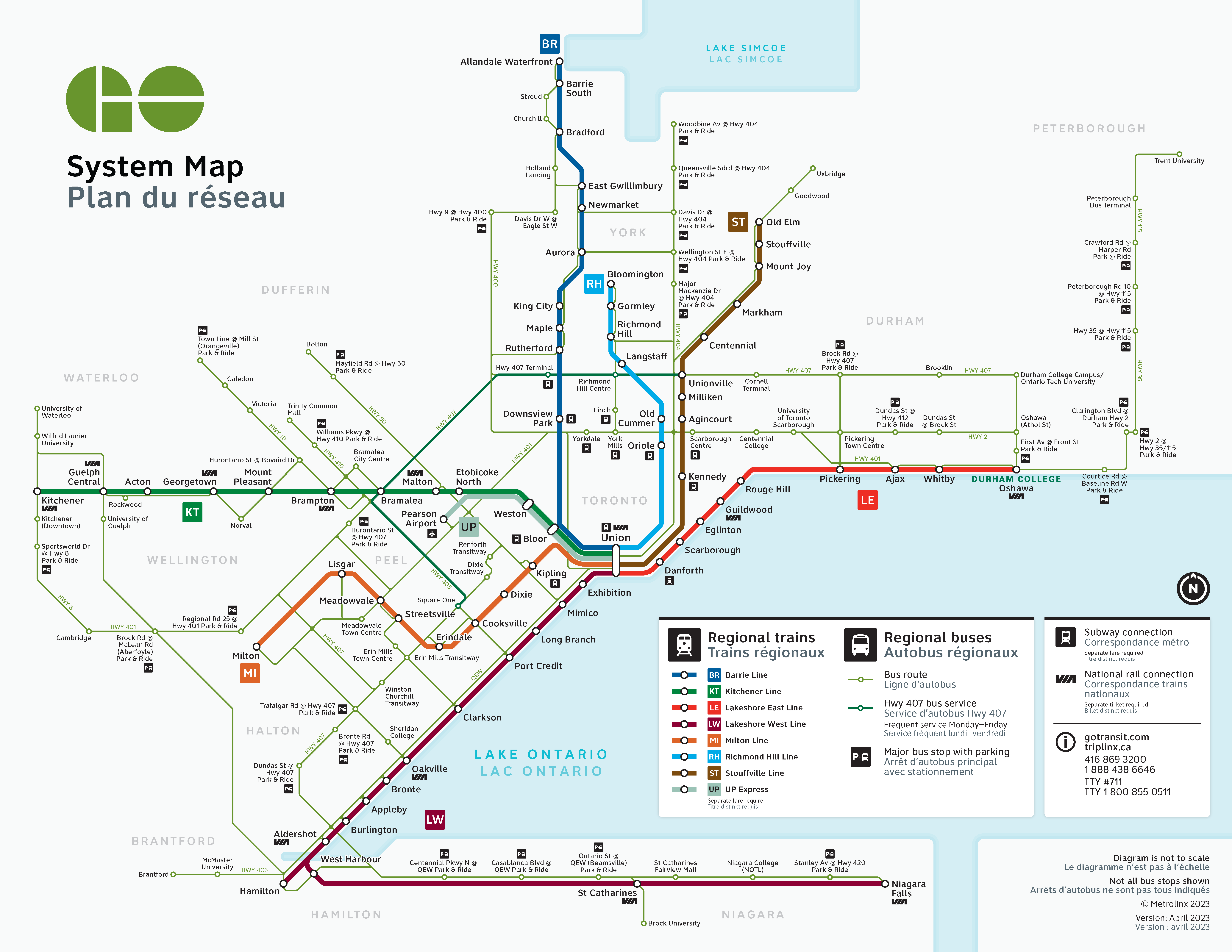}
  \caption{The system map of GO Transit, copied from GO Transit website\footnote{https://www.gotransit.com/en/system-map}}
  \label{fig:system_map}
\end{figure}

To exemplify the functionality of these applications, we have utilized information, including GTFS information, tweets, system alerts, etc, from GO Transit's official website\footnote{https://www.gotransit.com/en} and Twitter\footnote{https://twitter.com/GOtransit}. GO Transit serves as the primary regional public transit service provider in the Greater Toronto and Hamilton Area, catering to an extensive ridership of over 70 million passengers annually. As shown in Figure~\ref{fig:system_map}, the GO Transit network includes seven commuter rail lines supplemented by several commuter bus lines.

The three applications are built based on the following environment and tools:
\begin{itemize}
    \item \textbf{System}: Mac OS 14.0
    \item \textbf{Processor}: 2.4 GHz 8-Core Intel Core i9
    \item \textbf{Programming Language}: Python
    \item \textbf{UI framework}: Streamlit
    \item \textbf{Large Language Model}: GPT-3.5-Turbo
    \item \textbf{Prompt Framework}: LangChain
\end{itemize}
\subsubsection{Tweet Writer}
Contemporary transit agencies employ various forms of service alerts, encompassing on-vehicle or on-station displays (CleverVision~\footnote{https://www.cleverdevices.com/products/clevervision/}), official websites (TTC~\footnote{https://www.ttc.ca/service-alerts}) or mobile applications (Transit~\footnote{https://resources.transitapp.com/}), and text/SMS or email notifications (Peterborough Transit~\footnote{https://www.peterborough.ca/en/news/service-alerts.aspx}). Additionally, there is an observed trend where transit agencies increasingly use social media platforms for disseminating service updates. However, it is noteworthy that the generation and maintenance of official social media pages entail significant human effort. Furthermore, the issuance of service alerts often encounters delays between the actual service changes and their communication to the customer side.

The first application we present is the "Tweet Writer", which automates the generation of tweet content for real-time trip alerts or station alerts in the transit system. This application serves to reduce the workload of the staff in the media department of the public transit agency and assist them in providing timely information updates to customers.

\begin{figure}[!htbp]
  \centering
  \includegraphics[width=0.65\textwidth]{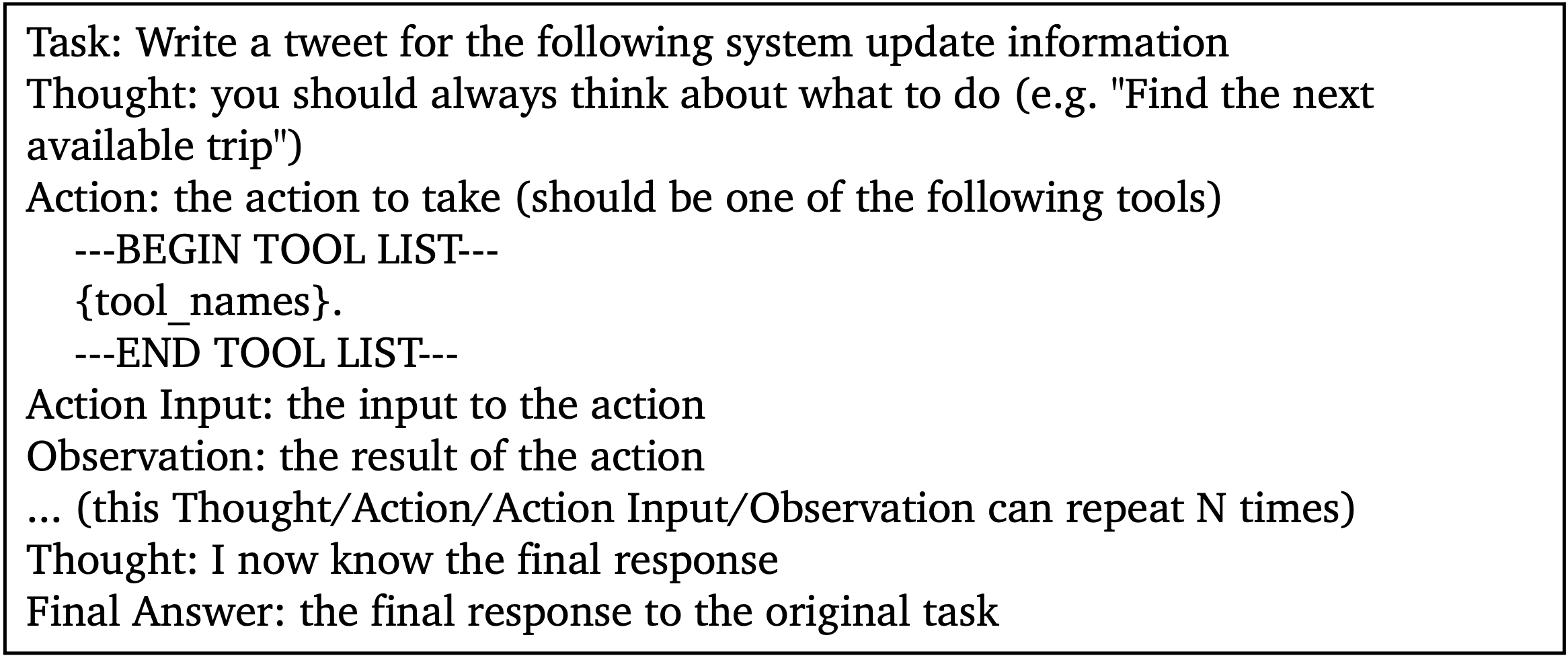}
  \caption{The prompt segment for LLM that uses the idea of CoT in Tweet Writer application}\label{fig:cot_tw}
\end{figure}

In Tweet Writer, the integration of CoT, as shown in Figure~\ref{fig:cot_tw},
empowers the LLM to adopt a specified thinking manner. This enables the model to automatically assess the task, utilize tools to interact with the resource database and explore valuable supplementary information guided by CoT prompts that are already integrated within the system. As a result, the LLM achieves a high level of automation in generating system alert tweets with accurate information. It is essential to emphasize that this tool is envisioned as an assistant to the communication staff rather than replacing their role in the process.

\begin{figure}[!htbp]
  \centering
  \includegraphics[width=0.85\textwidth]{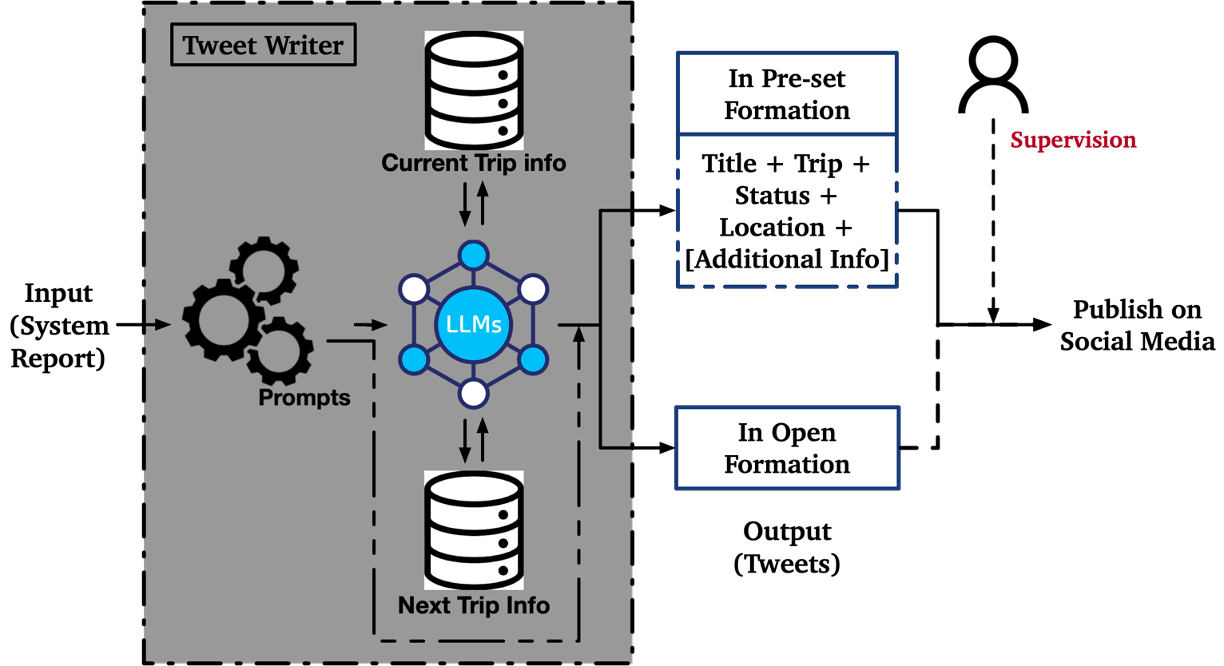}
  \caption{Workflow for Tweet Writer}\label{fig:tws}
\end{figure}

As illustrated in Figure~\ref{fig:gp} and~\ref{fig:tws}, the primary inputs for this application comprise real-time service alerts or system log information, which contains updates on trip status, service changes, and disruptions affecting commuters' journeys. These inputs are sourced through the front-end layer of the application, which is connected to the route source of real-time service alerts or open data portals where such alerts are published in GTFS-RT format. In the information extraction layer, the LLM seamlessly interacts with the database API through the "Thought/Action/Action Input/Observation" loop, as depicted in Figure~\ref{fig:cot_tw}. Depending on the type of alert, two main tasks can occur: finding trip information and determining the next available trip. To facilitate these tasks, we provide the LLM with specific functions: the trip information finding function and the next available trip finding function. Each function is accompanied by a concise instruction, detailing its purpose, required inputs, and the context in which the LLM should utilize it. The instructions and function names are given to the LLM in the prompt. The LLM only needs to mention the name of the function and input in their thought using natural language, and there is an additional function extracting the function name and input information, and running the corresponding function on the local computer.

Furthermore, within each computer-aided function, there is an LLM component responsible for parsing input into structured information, enabling rational database queries through the API function. This approach allows the main LLM to call aided functions with ease, as it only needs to provide sufficient information in its queries in natural language. The LLM inside the function will extract the useful information in the input into a specific formation, and then run the query API. Once all the relevant information is returned from the rational database, the LLM concludes the CoT loop and generates and returns the corresponding tweet to the front-end layer.

\begin{figure}[!htbp]
  \centering
  \includegraphics[width=0.9\textwidth]{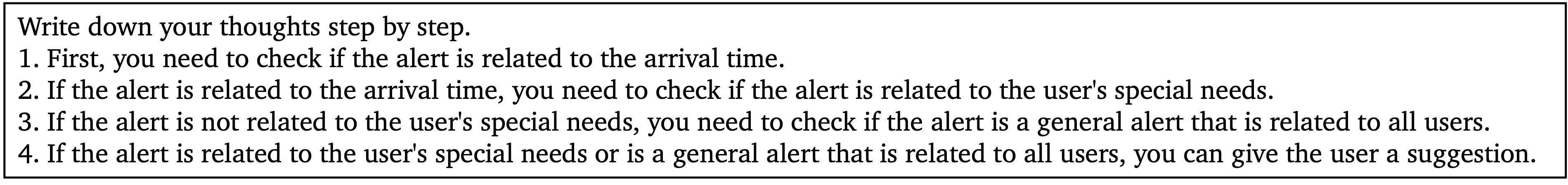}
  \caption{The prompt segment for LLM in Tweet Writer, indicating the specific formation we want it to follow}\label{fig:twsf}
\end{figure}

\begin{figure}[!htbp]
  \centering
  \includegraphics[width=0.8\textwidth]{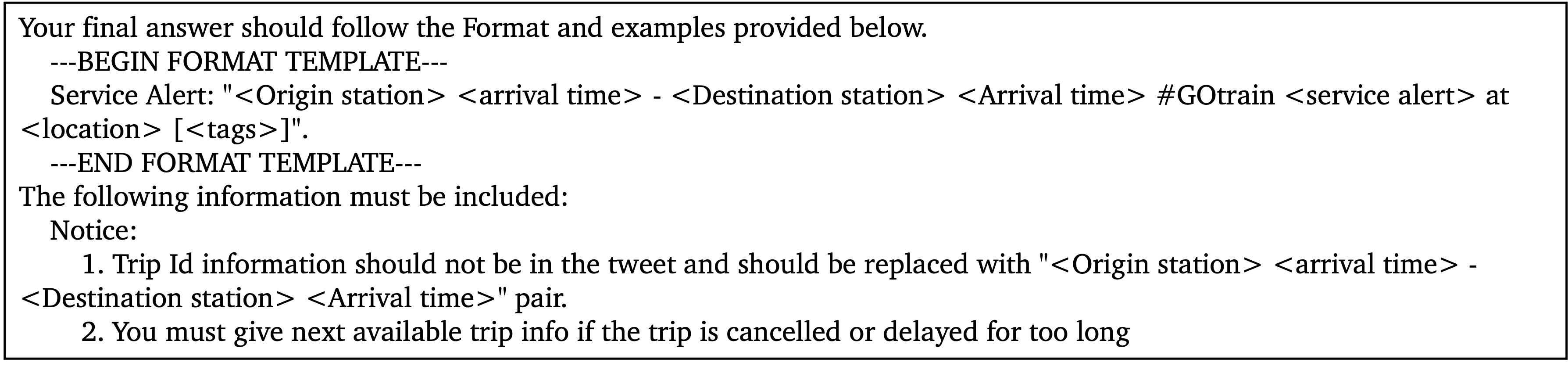}
  \caption{The prompt segment for LLM in Tweet Writer, indicating it can use any format}\label{fig:twop}
\end{figure}

Through this process, the LLM effectively transforms hard-to-interpret information into an intuitive format for customers. Additionally, the model offers trip suggestions when trip status deviates from the norm. The LLM's flexibility allows us to specify a desired tweet format for consistency, or grant more freedom to the LLM to generate content akin to that of a human or influencer, enhancing communication and making it more engaging and relatable to the general public. In Figure~\ref{fig:twsf}, and Figure~\ref{fig:twop}, we demonstrate the prompt segment that guides the LLM in the application to generate tweets either following a specific format or in an open format.

\begin{figure}[!htbp]
  \centering
  \includegraphics[width=0.8\textwidth]{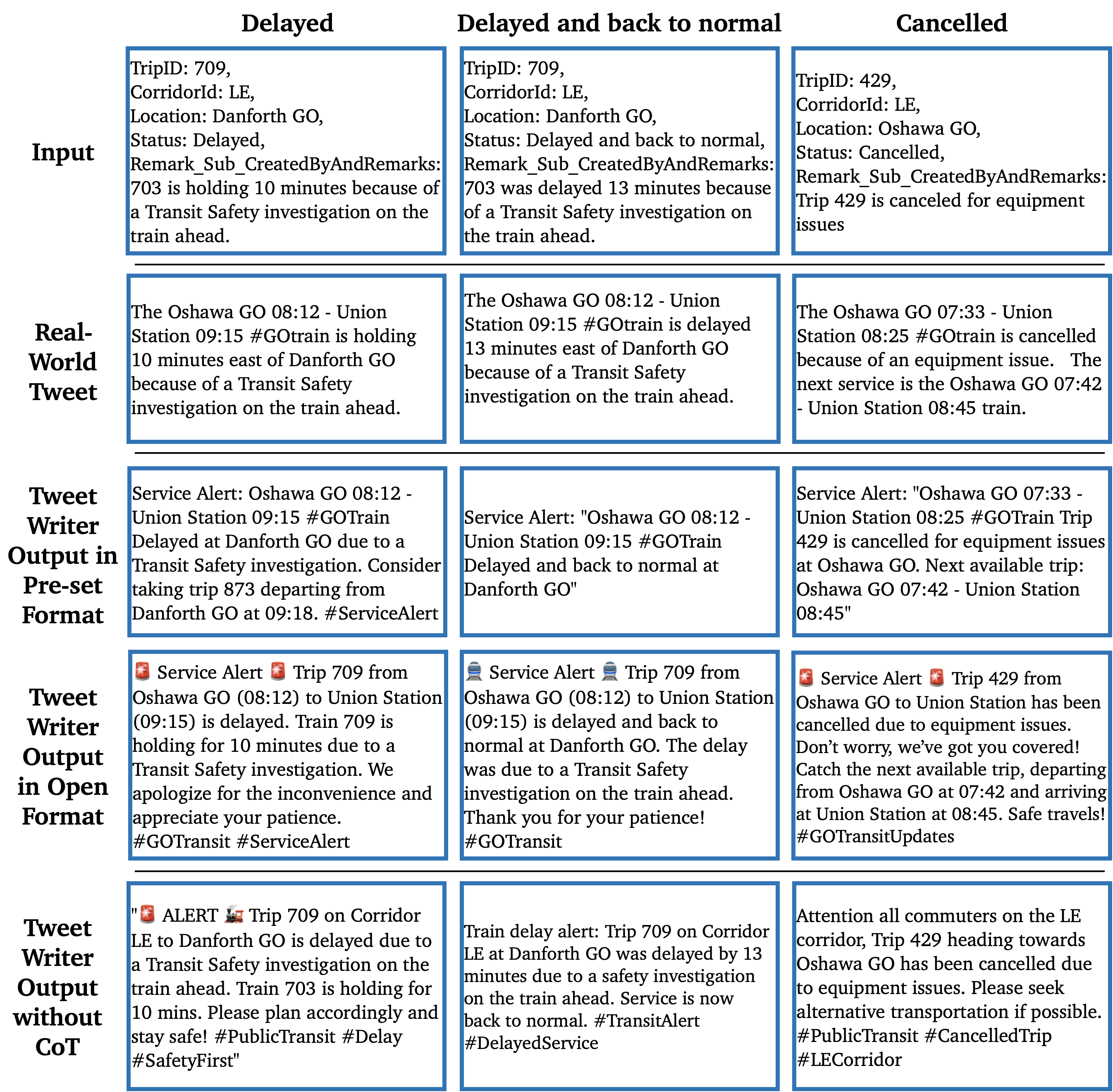}
  \caption{Demonstration of the outputs from Tweet Writer for three different trip statuses: delayed (on hold), back to move, and canceled.}\label{fig:tw}
\end{figure}

Figure~\ref{fig:tw} presents example outputs from the application, showcasing three different types of trip alerts: delayed (on hold), back to move, and canceled, along with two tweet formats: pre-set format and open format. These trip alerts are derived from real-world data within GO Transit's transit system, and we provide corresponding real-world tweets written by humans for each alert. Upon observation, the original alerts lack specific trip details, such as origin, destination, departure time, and arrival time. However, the LLM demonstrates its capability to automatically query the required information from the database and include it in the tweets. Additionally, the model autonomously searches for the next available trip to present to customers. Moreover, the tweets in the open format are embellished with emojis and comforting information. In comparison to human-written tweets, the application excels in providing tweets with similar essential information, while rendering them more captivating and enjoyable to read. 

Additionally, as part of our demonstration, we've evaluated the effectiveness of CoT, as described in~\ref{fig:cot_tw}. In Figure~\ref{fig:tw}, we've included the output of the application with simple prompts, where the prompts instruct the LLM to generate tweets based on input system alerts. However, the generated tweets vary significantly in their format and content. Specifically, the generated tweets lack essential information such as the trip's origin, destination, and corresponding arrival time. Furthermore, the hashtags used in the tweets lack consistency and fail to indicate the service provider, in this case, GO Transit. Meanwhile, the LLM fails to provide alternative trips for customers when the trips are canceled or delayed. These inconsistencies and omissions can confuse readers and hinder their ability to efficiently extract useful information. Moreover, tweets generated without the guidance of CoT may contain inaccuracies. For instance, a tweet about a delay situation (the one on the left-bottom in Figure~\ref{fig:tw}) incorrectly stated the trip's destination as Danforth Go, even though the train was already at the Danforth Go station. These observations underscore the importance of utilizing CoT to construct prompts. Doing so guides the LLM within the application, resulting in the generation of tweets that are consistent, information-rich, and accurate, ultimately enhancing the user experience and the efficiency of information retrieval. The proposed application is expected to streamline and enhance the communication staff's ability to post timely and informative tweets about the system's status, ultimately leading to increased efficiency. It is essential to emphasize that this tool is envisioned as an assistant to the communication staff rather than replacing their role in the process.

The proposed application is expected to streamline and enhance the communication staff's ability to post timely and informative tweets about the system's status, ultimately leading to increased efficiency. It is essential to emphasize that this tool is envisioned as an assistant to the communication staff rather than replacing their role in the process.

\subsubsection{Trip Advisor}
Transit agencies typically provide a web page for customers to plan their public transit journeys. In Ontario, the two main trip planner service providers are Google Maps and TripLinx~\footnote{https://www.triplinx.ca/}, the official planner for the Greater Toronto and Hamilton Area (GTHA). Users input their origin, destination, and departure time. TripLinx offers additional options like walking distance and service provider preferences, but the form requires users to input all details at once, potentially complicating the user experience. Notably, the planner does not directly provide system alerts related to customers' trips, limiting its functionality.

The "Trip Advisor" application is designed to provide a more personalized experience for users when planning their trips. Compared with classical trip planners, as mentioned previously, our application allows users to engage in a more natural interaction without worrying about the form of the query. Additionally, the application takes into account individual user preferences and limitations, resulting in personalized trip options that prioritize a user-centric experience.

This application is an example application that shows how the application can query a structured database with unstructured input and generate output in natural language. The developed application is currently designed for desktop use; however, it can be readily adapted to function as a mobile application.

\begin{figure}[!htbp]
  \centering
  \includegraphics[width=0.8\textwidth]{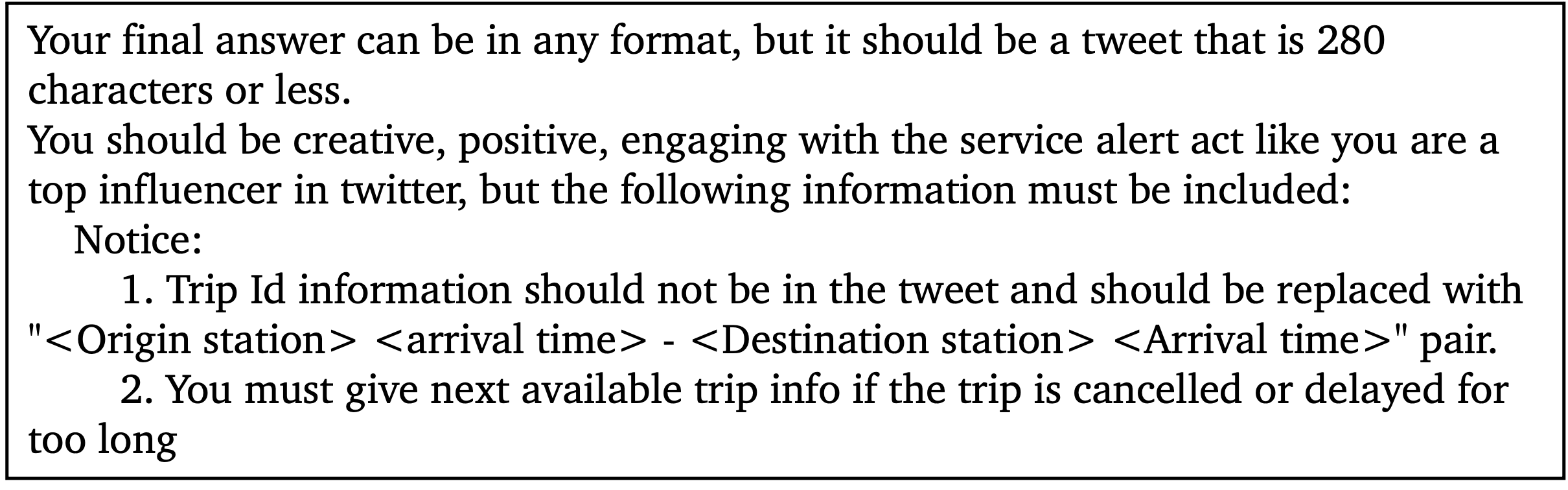}
  \caption{The prompt segment for LLM that uses the idea of CoT in Travel Advisor application}\label{fig:cot_ta}
\end{figure}

In the Trip Advisor application, we leverage the CoT framework to guide the LLM in determining the relevance of transit system alerts to individual passengers, as shown in Figure~\ref{fig:cot_ta}.
The LLM systematically follows predefined steps to assess the relationship between the system alert and the passenger's specific situation. It checks for any overlap or alignment between the alert content and the passenger's needs before making personalized trip suggestions.

The key inputs for this application include Origin-Destination (OD) information, travel time preferences, and any specific requirements or special needs the user may have. These inputs are provided through direct user interaction, typically through conversational dialogue or specific user queries. Based on the user inputs, the application provides two main outputs: customized trip suggestions and customized trip alerts. The trip suggestions are tailored to the individual's preferences and the trip alerts cover any relevant changes or updates that could affect their planned journey. In this application, the LLM helps parse and understand the user input, extracting key details like the desired OD, travel time, and special needs. Additionally, the application possesses the ability to comprehend the user's sentiments and emotions, determine whether the conversation should continue or be concluded, and execute subsequent steps, such as trip information searching and trip suggestion generation.

\begin{figure}[!ht]
  \centering
  \includegraphics[width=0.45\textwidth]{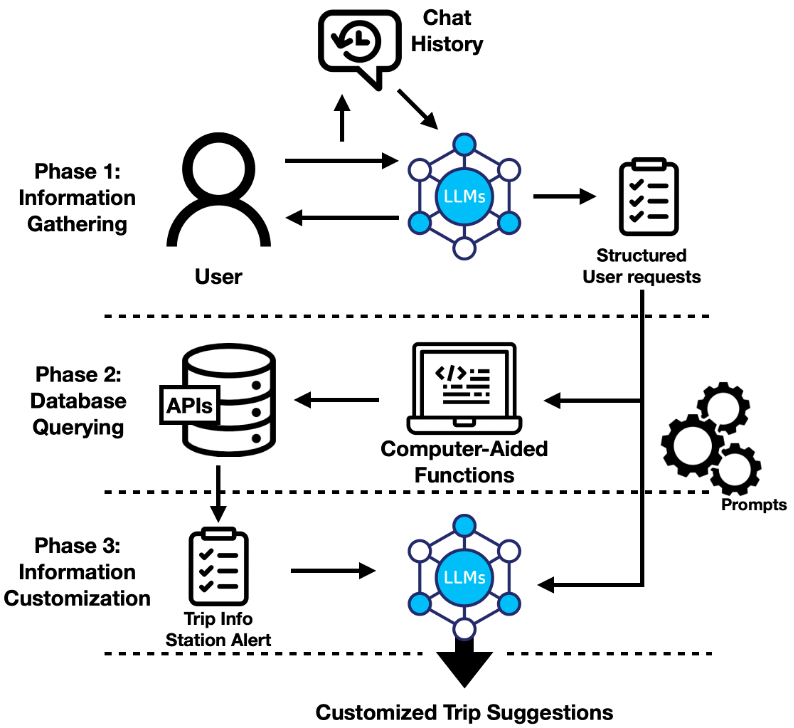}
  \caption{Demonstration of the workflow for Trip Advisor}\label{fig:ts}
\end{figure}

\begin{figure}[!htbp]
  \centering
  \includegraphics[width=0.6\textwidth]{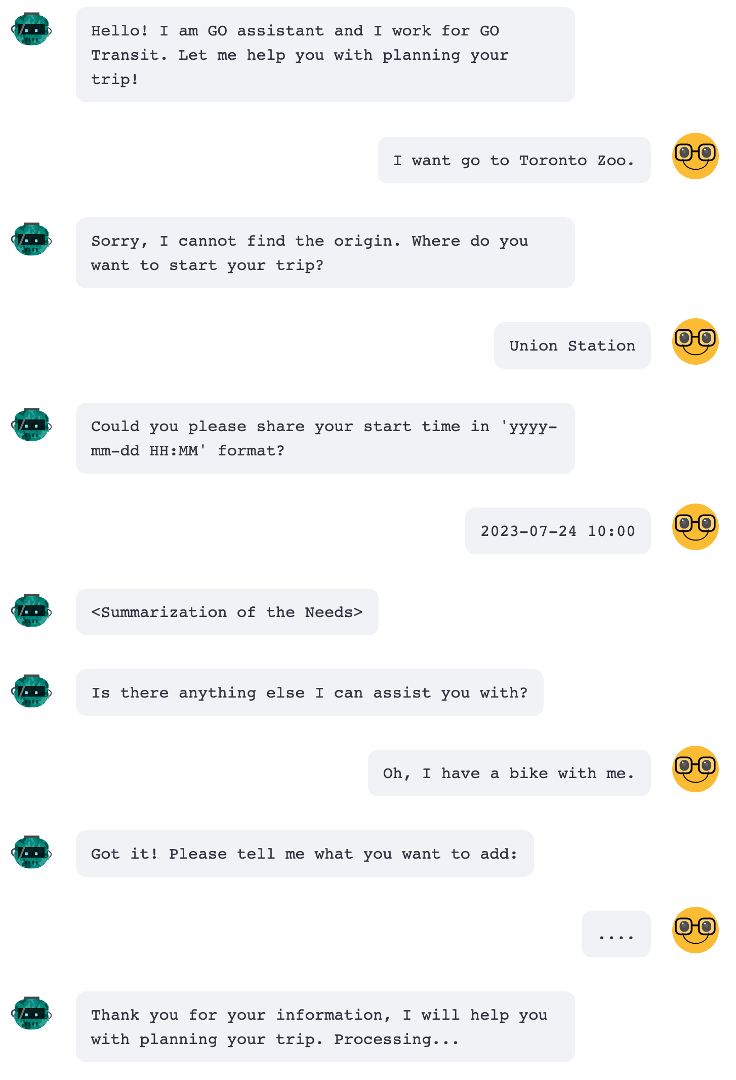}
  \caption{Example conversation between the application and a user}\label{fig:tse}
\end{figure}

\begin{figure}[t]
  \centering
  \includegraphics[width=0.65\textwidth]{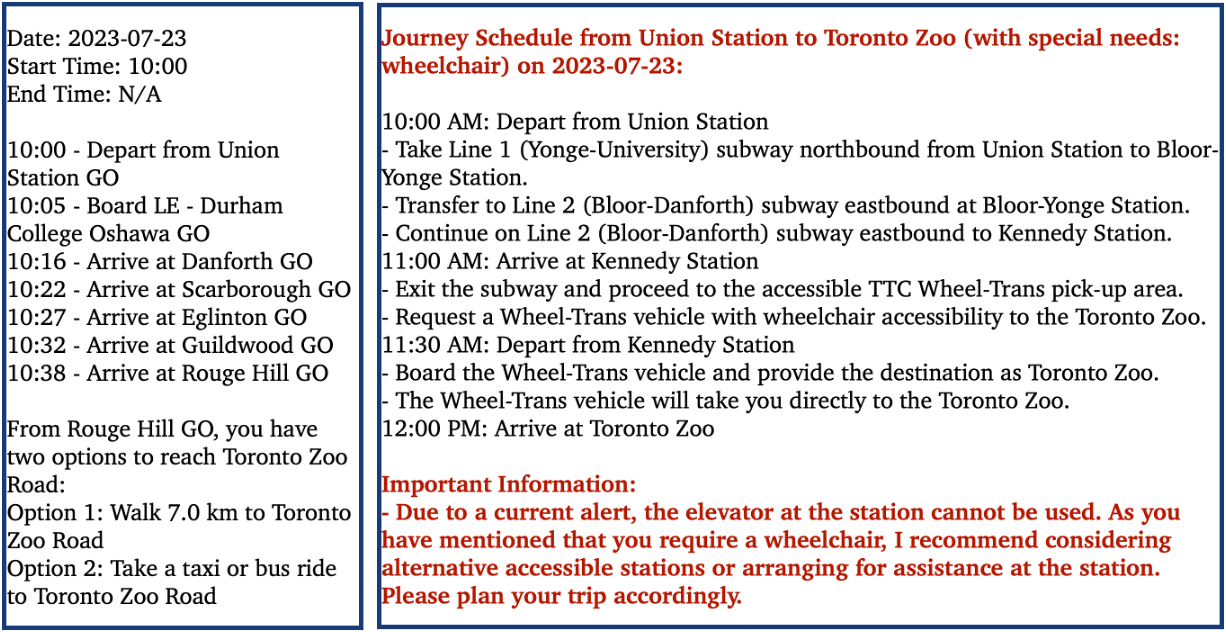}
  \caption{Trip suggestions made by the application. The picture on the left is the suggestion where the user has no special needs, while the picture on the right side is the suggestion made for a user using a wheelchair.}\label{fig:tsee}
\end{figure}

As depicted in Figure~\ref{fig:ts}, the application follows three main phases. In the first phase, "Information Gathering," essential information is collected through a conversation between the user and the application. With the memory mechanism we have implemented for the LLM, users are not required to provide all the information at once. The LLM can continuously engage with the user, requesting additional details if the current information is insufficient to make a decision. The LLM also attempts to identify any special needs expressed by the user. Once all the necessary information has been gathered and the LLM detects that the user does not wish to continue the conversation, it generates structured information in JSON format based on the chat history. This information includes the origin-destination details, departure time, and any special needs specified. This information is used to query a rational database through the functions we provided to the LLM, which are built upon the Database API and can give LLM access to search for trips with origin, destination, and departure time information. Unlike the Tweet Writer application, where additional LLM is embedded inside the computer-aided function, the input construction step is performed by the main LLM in the application workflow. The reason is that the structured information serves different phases, and processing it with a single LLM enhances application efficiency. The structured information is aligned with the input parameter requirement of the query functions. The retrieved information, along with the original user information collected in Phase 1, is utilized by the LLM in Phase 3 to generate personalized trip suggestions. An example conversation between a user and the application is presented in Figure~\ref{fig:tse},
and an example output from the application is shown in Figure~\ref{fig:tsee}.
As we can see, depending on different needs or requirements, the application can provide customized trip suggestions for the user.

\subsubsection{Policy Navigator}
\begin{figure}[!ht]
  \centering
  \includegraphics[width=0.8\textwidth]{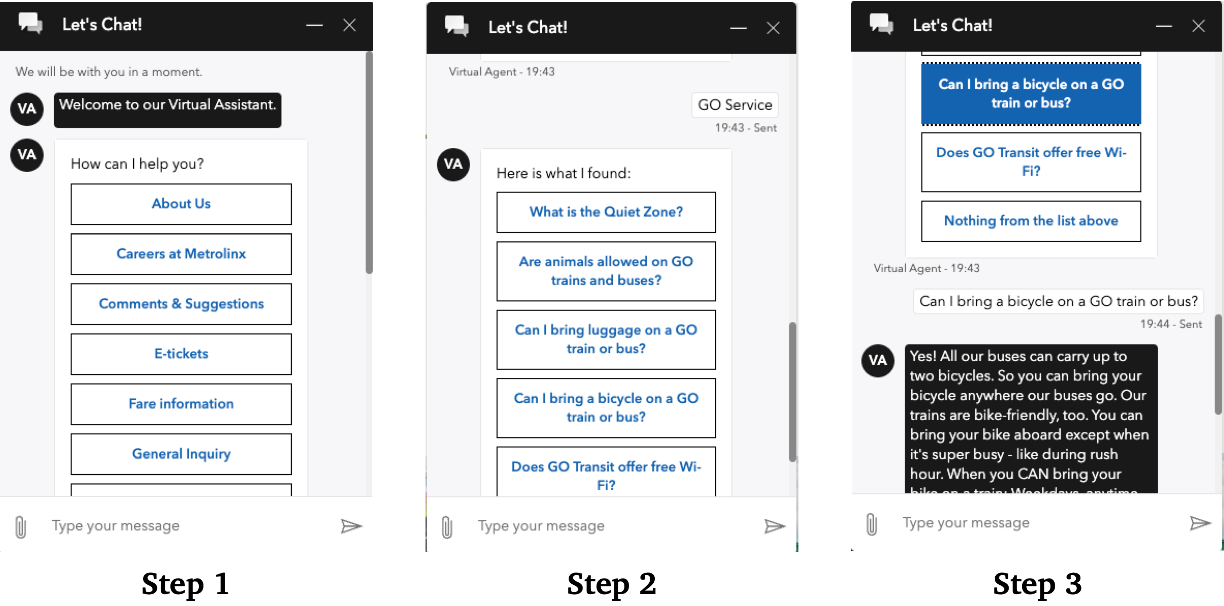}
  \caption{Process we have to go through when querying whether we can bring a bike on the GO train. The screenshots are taken from GO Transit Website}\label{fig:pnp}
\end{figure}

\begin{figure}[!ht]
  \centering
  \includegraphics[width=0.46\textwidth]{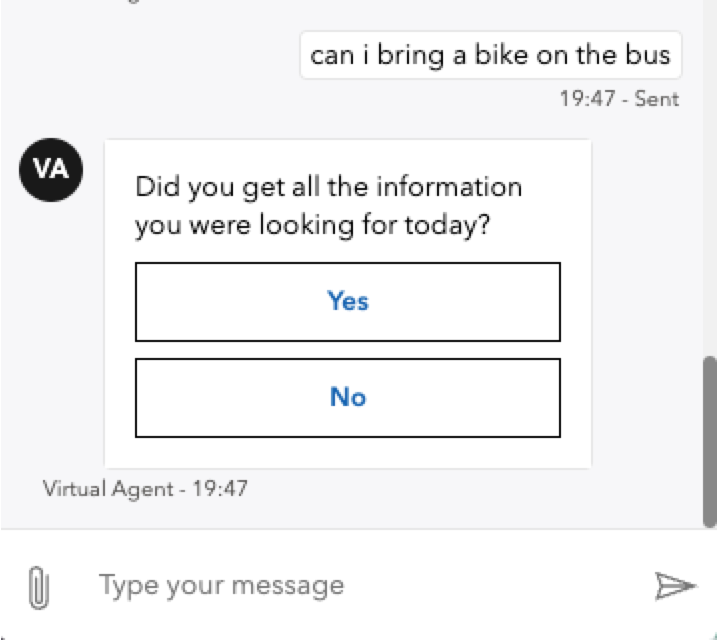}
  \caption{Ask directly to the virtual assistant. The screenshots are taken from GO Transit Website}\label{fig:pnp2}
\end{figure}

The "Policy Navigator" application utilizes LLM to provide personalized policy advice in response to user queries. First, we examine the Virtual Assistant\footnote{https://www.gotransit.com/en} deployed on the GO Transit website as an example of current tools that provide information to users. As shown in Figure~\ref{fig:pnp}, 
the GO's virtual assistant requires users to navigate through a series of steps, select a question class, and choose from a list of candidate questions under that selected class. The retrieved information often includes irrelevant details. When users attempt to ask a question directly to the virtual assistant, the system could return no response as the system's pre-defined questions were not written in the same format as the users' queries, as demonstrated in Figure~\ref{fig:pnp2}.

\begin{figure}[!ht]
  \centering
  \includegraphics[width=0.7\textwidth]{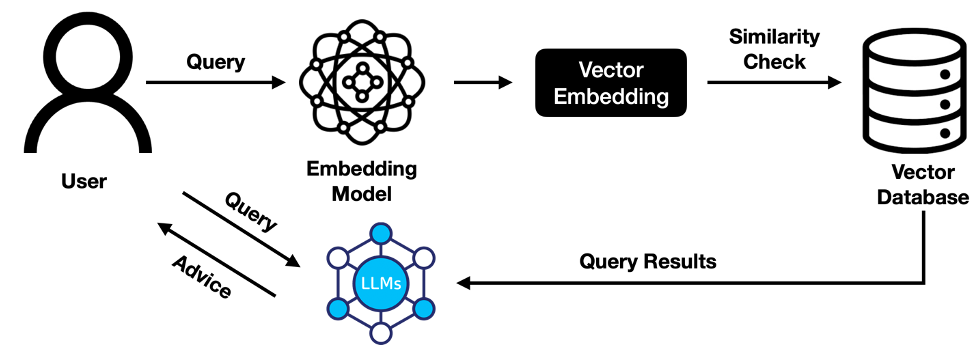}
  \caption{The workflow of Policy Navigator}\label{fig:pn}
\end{figure}

The current policy assistant applications provided by public transit agencies often resemble navigation bars, resulting in redundant user interactions and the retrieval of excessive information. To address these limitations, our application accepts user queries formulated in natural language. In response to these queries, the Policy Navigator generates customized policy advice that directly addresses the user's query. The application aims to guide users to the most relevant policy documents or specific sections for further reading. The workflow of the Policy Navigator is illustrated in Figure~\ref{fig:pn}, 
with the application utilizing the policy documents from GO Transit as its resource base, which will be stored in a vector database. The application retrieves the policy segment that is most relevant to the user's question and refines the feedback based on the retrieved information and the user's query.

\begin{figure}[!ht]
  \centering
  \includegraphics[width=0.85\textwidth]{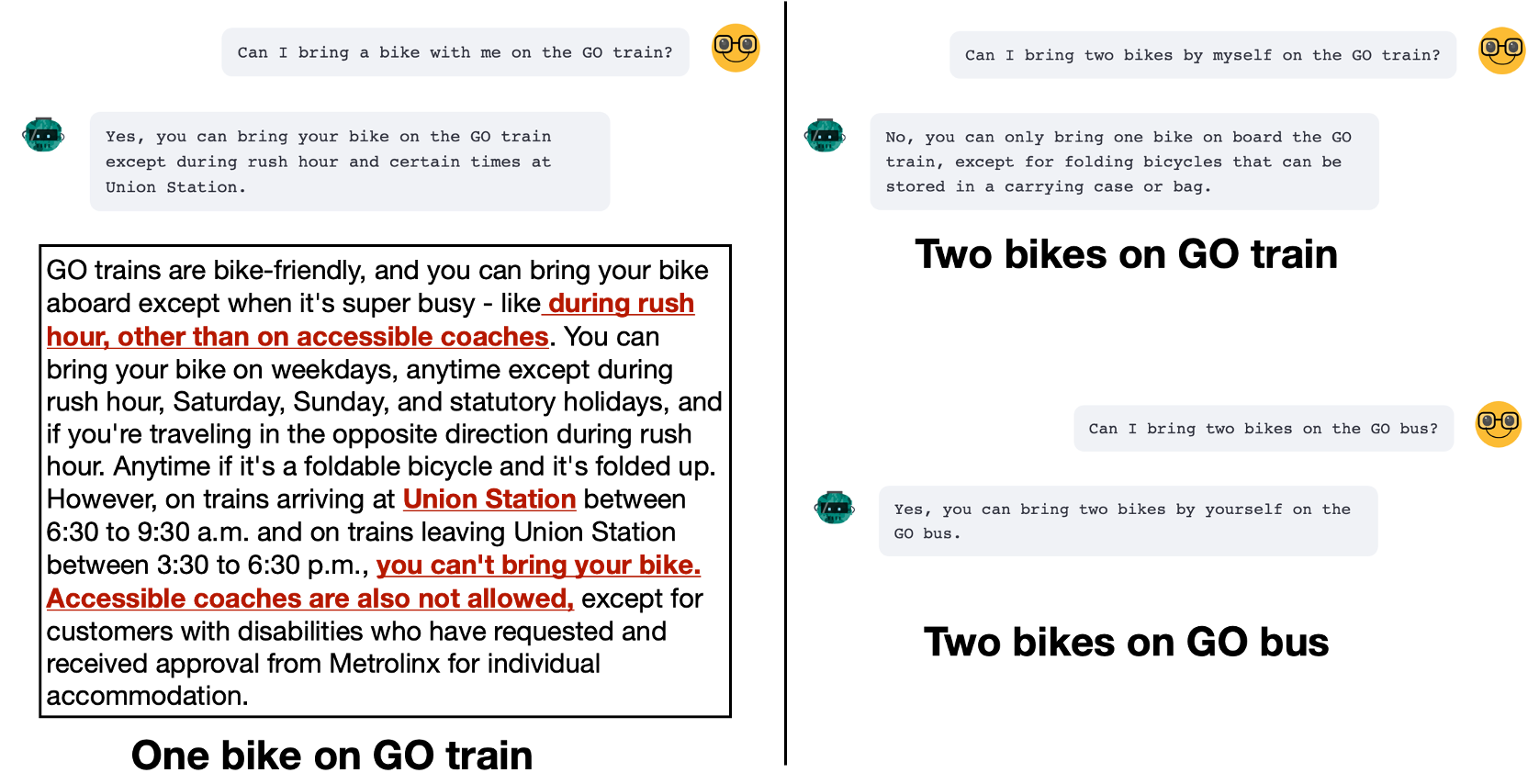}
  \caption{Example answers from Policy Navigator for answering three slightly different questions related to the carry-on bike policy in GO Transit}\label{fig:pnn}
\end{figure}

As shown in Figure~\ref{fig:pnn}, 
the feedback provided by the application varies with slight changes in the user's query, demonstrating its ability to provide direct and clear answers that meet the users' needs. Additionally, users have the option to request the original retrieved information as supplementary material to ensure the reliability of the results.

\begin{figure}[!ht]
  \centering
  \includegraphics[width=0.4\textwidth]{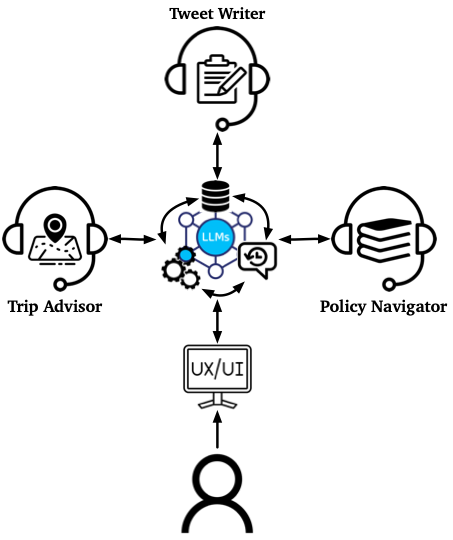}
  \caption{The structure of TransitTalk Digital Assistant}\label{fig:fw}
\end{figure}

Although the applications are currently implemented as separate entities, their seamless integration into the overarching framework, known as "TransitTalk Digital Assistant," is a straightforward process. This integration transforms them into a unified whole, providing users with an uninterrupted and cohesive experience when seeking information related to trip details, station alerts, and policy inquiries. As shown in Figure~\ref{fig:fw}, by incorporating the memory mechanism into this comprehensive framework, the system gains the capabilities of cross-application information sharing and the retention of user preferences and limitations. Driven by the capabilities of the LLM, users are free from the concern of which specific application they are interacting with. The chat history becomes shared within the three applications, enabling the LLM to intelligently select the appropriate application to retrieve the required information. For instance, a user can inquire about trip specifics, such as traveling from point A to B with a bicycle. Concurrently, queries related to relevant policies, such as carrying bicycles on transit, can be recognized and addressed. After Trip Advisor generates trip information, related station alerts along the trip are transformed into user-friendly tweets using the Tweet Writer application. On the other hand, the user can start by asking policy-related questions, and when he/she transitions to a trip query, they do not need to provide the same background information.

\section{Discussion} \label{sc:discussion}
\subsection{Limitations and Challenges Using LLM}
Although LLMs are powerful tools for information communication between computers and humans, they have certain limitations. One of the main limitations is their lack of understanding of that which they do not know. This includes their knowledge boundaries, areas of expertise, the accuracy of their answers, and how to correct their mistakes. These limitations can result in incorrect answers that may be biased, fabricated, or contain reasoning errors. Additionally, LLMs may have limitations due to outdated knowledge in their training datasets.

These limitations pose several challenges when developing applications for the public transit system. Firstly, ensuring the reliability of the system becomes difficult, including the trustworthiness and stability of the results. Since the models are not specifically trained on domain-specific knowledge, their professionalism in handling complex decision-making tasks is questionable. Moreover, evaluating the output from the model becomes challenging, as there are no benchmarks for using LLMs in the public transit system. It is difficult to assess the correctness and reasonableness of natural language answers.

Furthermore, integrating LLMs into applications introduces flexibility and creativity, as they can make decisions in various circumstances without explicitly enumerating all possible interactions between the application, the user, and the database. However, this flexibility and creativity come at the cost of efficiency and consistency. LLMs may get stuck in thought loops and struggle to adhere to the given requirements, resulting in a lack of imperative consistency.

\subsection{Best Practices Using LLM in the Public Transit System}
Several best practices can be employed to address the aforementioned challenges.

Firstly, by using appropriate prompts one can enhance the reliability of the LLM's output. By providing specific instructions in the prompts, such as the desired problem-solving approach, format, tone, and references, we can guide the LLM toward generating more reliable answers. Open-source tools like LangChain~\cite{longchain} can facilitate the development of effective prompts. Furthermore, there are also AI tools\footnote{https://gptforwork.com/tools/prompt-generator}$^,$\footnote{https://huggingface.co/merve/chatgpt-prompt-generator-v12} that can help generate proper prompts, if the user can specify their tasks.

To improve professionalism, fine-tuning techniques can be utilized to tailor the LLM with domain-specific knowledge. This approach allows us to enhance the LLM's expertise in a particular area, leveraging relatively fewer training resources and time.

For evaluation purposes, a higher-version LLM model, such as GPT-4, can be employed as a critic to assess the quality of the application's output or the similarity between the generated results and the expected answers.

However, during the development of the three transit applications described earlier, we realized the importance of treating the LLM as a tool with limited intelligence. The public transit domain demands a high level of reliability, as unreliable answers can erode customer trust. While the LLM serves as a valuable tool for information conversion, it is crucial to regularly check its results using computer-aided functions or human supervision to ensure the accuracy of the application's output. For instance, in Tweets Writer, the tweets generated by the AI should undergo a human review within the transit agency before being published on social media. This process ensures that human workers have access to both the generated tweets and the raw information generated by the system, maintaining the accuracy and appropriateness of the content shared with the public. Similarly, in the workflow of Trip Advisor, the AI is programmed to request basic information from users, such as origin, destination, and preferred time. The flexibility of the LLM allows for variations in when and how these questions are posed. However, to ensure the integrity of the LLM's output, the questions must be either explicitly asked or inferred from the customer's chat history. This careful approach guarantees the reliability and relevance of the information provided by the LLM in generating personalized transit trip suggestions, even though it may restrict the working space of LLMs and require additional coding or human effort. 

\section{Conclusions and Future Work} \label{sc:con}
The technical guidance offered in this study is not solely tailored to the proposed three applications; rather, it serves as a comprehensive roadmap for transit researchers and practitioners. Through the exploration of the three showcased applications, we have addressed diverse information transformation processes prevalent in transit systems: the conversion from structured information to neural language information, from neural language information to structured information, and the transformation between neural language information types. This study elucidates how these transformations can be effectively executed using LLM.

Transit agencies can benefit from the insights garnered in this research. Firstly, the proposed framework offers a systematic approach to incorporating LLMs into various aspects of transit operations. By embracing this methodology, transit agencies can streamline communication processes, automate service alerts, and deliver tailored responses to customer inquiries. The showcased applications serve as practical models, illustrating how LLMs can be leveraged to improve the efficiency of system updates, personalized trip planning, and policy-related interactions. While these applications serve as early-stage showcases, they lay the groundwork for further refinement and adaptation by real-world transit agencies. The demonstrated potential for improvement underscores the scalability and practicality of integrating LLMs into transit operations, marking a transformative stride towards more efficient and responsive customer service systems in the public transit domain.

There are several promising avenues for future research and development in leveraging LLMs in public transit systems. These directions aim to enhance customer service, improve the efficiency of automatic reporting systems, and advance the capabilities of transit control assistant systems. Firstly, further refinement of the chatbot applications discussed in this paper is needed to provide even more accurate and contextually appropriate responses to customers. Secondly, improving the working experience of transit system staff can be achieved through the implementation of an automatic reporting system. Lastly, enhancing transit control assistant systems is an important area of future development. By leveraging Cross-Modal LLMs, real-time transit situations can be understood and analyzed, enabling the generation of control suggestions for transit control centers. Fine-tuning LLMs with domain-specific information, such as historical control records, control strategies, or guidelines, can further enhance the effectiveness and efficiency of control recommendations.

\section{Acknowledgement}
The core of the three applications presented in this paper is powered by GPT-3.5-Turbo, performing natural language understanding, information extraction, and natural language generation.

ChatGPT is employed for paper proofreading and grammar checking.

It is important to note that no LLM was utilized in generating the content of this paper.

\bibliographystyle{plainnat}
\bibliography{reference}

\end{document}